# Modeling the directed transmission and reflection enhancements of the lasing spaser in active metamaterials

Zheng-Gao Dong,[1,*] Hui Liu,[2,†] Tao Li,[2] Zhi-Hong Zhu,[2] Shu-Ming Wang,[2] Jing-Xiao Cao,[2] Shi-Ning Zhu,[2] and X. Zhang[3]

[1]*Physics Department, Southeast University, Nanjing 211189, China*

[2]*National Laboratory of Solid State Microstructures, Nanjing University, Nanjing 210093, China*

[3]*25130 Etcheverry Hall, Nanoscale Science and Engineering Center, University of California, Berkeley, California 94720-1740, USA*

[*]Electronic address: zgdong@seu.edu.cn

[†]Electronic address: liuhui@nju.edu.cn;

URL: http://dsl.nju.edu.cn/dslweb/images/ plasmonics-MPP.htm

Simultaneously enhanced reflectance and transmittance greater than 35 dB are demonstrated for the lasing spaser (or spasing) behavior in an active fishnet metamaterial. In mimicking a lasing cavity, an equivalent active slab with Lorentz dispersion for the index of refraction is established to model the spasing metamaterial through the Fabry-Perot effect. Numerical and theoretical results show good agreement in the equal enhancement of reflectance and transmittance, as well as the non-monotonic dependence of the spasing intensity on the gain coefficient. In addition, directed emission of the spasing beam is verified numerically.

## I. INTRODUCTION

Metamaterial, as a kind of artificial material, offers a versatile application potential from microwave to visible spectrum. For example, a left-handed metamaterial can bend the light beam negatively,[1,2] a super-lensing metamaterial can beat the diffraction limit and reconstruct the object in the imaging domain with a subwavelength resolution,[3,4] while a cloaking metamaterial can realize a novel invisibility device by forcing electromagnetic waves to propagate along its surface and leave it in original directions.[5,6] Recently, a spasing metamaterial (i.e., lasing spaser) was proposed by Zheludev *et al*. to realize a coherent stimulated emission with a giant enhancement in the transmitted waves,[7] which is very attractive and promising for the realization of a nanolaser.[8,9]

In a lasing spaser configuration,[7] a two-dimensional metallic metamaterial composed of asymmetrical split rings could transmit the incident light with orders of magnitude enhancement under the assistance of an active layer at moderate gain level. Intuitively, the secret of a spasing metamaterial, as compared with an ordinary gain process, lies in the surface plasmon resonance of the metallic metamaterial. On the one hand, plasmon resonance brings a strong plasmonic field localized in gaps of metallic elements. As will be discussed in section III, this contributes to the moderate gain requirement for great amplification in spasing metamaterials. On the other hand, we believe that diverse effective index of refraction of a metamaterial, which varies considerably around the resonance frequency, could take the role as a coated mirror does in a laser cavity to meet a certain transmission/reflection standard. More recently, by comparison between the metallic double-ring metamaterial and the asymmetrically split ring structure in ordered and disordered arrays,[10] it was argued that the coherence of the spasing could be guaranteed by synchronous oscillations of plasmonic currents in the ordered metallic array. In further, the dynamic response of the lasing spaser was studied by a toy model,[11] and a self-consistent calculation was developed to treat the combination system of gain-assisted metamaterials.[12] In this communication, an active fishnet metamaterial is investigated, both numerically and theoretically, to present the giant amplification of the reflected as well as the transmitted spasing behavior, to understand them from the viewpoint of the Fabry-Perot model, and to evaluate the emission directivity of the spasing behavior.

## II. FISHNET SPASING METAMATERIAL

Fishnet metamaterials, composed of alternating metal/dielectric layers with perforated holes, are currently popular in studying negative refraction, cloaking, and magnetic plasmon polaritions.[2,13-16] Figure 1 shows schematically the adopted two-dimensional fishnet metamaterial, assisted by an active interlayer, with $A_x = 496\,\text{nm}$, $A_y = 310\,\text{nm}$, and $L_x = L_y = 930\,\text{nm}$. The silver in fishnet structure is 62 nm thick and follows the Drude dispersion ($\omega_p = 1.37 \times 10^{16}\,\text{s}^{-1}$ and $\gamma = 8.5 \times 10^{13}\,\text{s}^{-1}$).[17,18] In addition, the interlayer host is polymethyl methacrylate (PMMA, optical index of refraction $n_0 = 1.49$) with 93 nm thickness, in which PbS semiconductor quantum dots are doped as the active inclusion. According to the emission property of PbS quantum dots, its gain coefficient is described in this work by a Gaussian distribution, with the maximum gain $\alpha_0$ at 1500 nm emission and a full width at half maximum (FWHM) 150 nm. As for the absolute value of the maximum gain $\alpha_0$ at 1500 nm emission, it is tunable and experimentally depends on the quantum dots density, pumping power, sample temperature, and so on.[19]

## III. RESULTS AND ANALYSIS

### A. Simultaneously enhanced transmission and reflection

By applying corresponding perfect electric and magnetic boundary conditions, a polarized plane wave with electric field component in the *x* direction and magnetic field component in the *y* direction is incident normally onto the fishnet metamaterial. Figure 2(a) shows the transmittance (black solid line) and reflectance (blue dashed line), numerically calculated by the full-wave finite element simulation in frequency domain.[20,21] It is found that the transmittance and reflectance are simultaneously amplified to an almost equal level greater than 30 dB for $\alpha_0 = -2.35 \times 10^3$ cm$^{-1}$ at the resonant frequency of 202 THz, and both of them decrease rapidly away from the resonant frequency (for different gain coefficients, larger spasing intensity is accompanied by smaller spasing bandwidth). On the other hand, the spasing intensity firstly increases and then decreases with the gain coefficient [Fig. 2(b)], that is, the giant enhancements in transmission and reflection only occur at a certain gain level. Generally speaking, the simulation results for the transmitted spasing are in agreement with the literature.[7] Nevertheless, to the best knowledge of ours, several issues about the spasing behavior in active metamaterials have not yet been interpreted systematically, such as the non-monotonic dependence of spasing intensity on the gain coefficient, the reflected spasing behavior, and the small gain requirement for giant enhancement.

As a matter of fact, the spaser concept was proposed somewhat in an attempt to mimic the laser mechanism.[22,23] The key difference is that for the former the amplification is realized via surface plasmon oscillation in metals, while for the latter it is light amplification and usually not involved in metals. In consideration of the similarity between the spaser and the laser, in the next section, we will try to model the spasing fishnet metamaterial in terms of a Fabry-Perot resonant cavity. Several issues regarding the spasing behavior in active metamaterials will be explained in terms of the Fabry-Perot effect.

### B. Fabry-Perot cavity model

Consider that the active fishnet metamaterial is equivalent to a flat slab with a complex index of refraction $\tilde{n} = n + i\kappa$, consequently, the reflectance and transmittance at the slab surfaces for a normal incident case are $R = [(n-1)^2 + \kappa^2]/[(n+1)^2 + \kappa^2]$ and $T = 1 - R = 4n/[(n+1)^2 + \kappa^2]$, respectively. Take into account the Fabry-Perot effect, as shown in Fig. 3(a), an incident light is at first partially transmitted into the slab, and then will be reflected back and forth between the two surfaces of the slab, meanwhile, with some proportion transmitted outside the slab for each reflection. The intensity summations of all the transmitted and reflected waves can be written as[24,25]

$$I_T = \frac{(1-R)^2 e^{-\alpha_{th} d}}{(1-Re^{-\alpha_{th} d})^2 + 4Re^{-\alpha_{th} d}\sin^2\frac{\varphi}{2}}$$

and $$I_R = \frac{R(1-e^{-\alpha_{th} d})^2 + 4Re^{-\alpha_{th} d}\sin^2\frac{\varphi}{2}}{(1-Re^{-\alpha_{th} d})^2 + 4Re^{-\alpha_{th} d}\sin^2\frac{\varphi}{2}},$$

where $\alpha_{th} = 4\pi\kappa/\lambda$ represents the theoretical gain coefficient of the slab and $\varphi = (4\pi/\lambda)nd$ is the phase difference after a round-trip transit. To characterize the active fishnet resonator, the real part of the index of refraction for the equivalent slab follows a Lorentz dispersion $n = n_0 - (f_p^2 - f_o^2)(f^2 - f_o^2)/[(f^2 - f_o^2)^2 + f_\gamma^2 f^2]$, where $f_p = 214.2$ THz, $f_o = 198.8$ THz, $f_\gamma = 10.86$ THz, and $n_0 = 1.49$ [black solid line in Fig. 3(b)]. The imaginary part of the index of refraction $\kappa = (\lambda/4\pi)\alpha_{th}$, with a Gaussian-profile gain coefficient $\alpha_{th} = -5.35\times 10^6 \times \exp[-(f-202.5)^2/8] + 4\times 10^5$, is shown by the blue dashed line in Fig. 3(b), where positive $\kappa$ represents greater loss in the metallic bilayer than gain in the active interlayer while negative $\kappa$ attributes to the opposite case. Additionally, the slab has a small thickness $d = 50$ nm, so that the interference fringes for $I_T$ and $I_R$ will be indistinguishable because of $d \ll \lambda$.

Before comparing the theoretical analysis with the simulation results in Fig. 2, a little discussion about the spasing condition, indicated by expressions of the intensity summations $I_T$ and $I_R$, would be helpful to understand the giant amplification characteristic of the Fabry-Perot spaser cavity. Obviously, the extremum of $I_T$ (also $I_R$) should occur at $\sin\frac{\varphi}{2} = 0$ and thus the round-trip amplification factor $f'$ is simplified to be $Re^{-\alpha d}$ [see Fig. 3(a)]. The summation expression $I_T$ as well as $I_R$ holds only if the gain inside the cavity does not overcompensate the output coupling and other possible losses, i.e., the factor $R\mathrm{e}^{-\alpha d}$ should not exceed 1. If $R\mathrm{e}^{-\alpha d} \geq 1$, which means the intensity inside the cavity is increased by the gain medium more than that is scattered outside, then in a mathematical sense both the transmitted and reflected spasing would be unrealistically infinite, and hence the expressions for $I_\mathrm{T}$ and $I_\mathrm{R}$ are not applicable for this case. In a realistic spaser configuration, to obtain the maximum spasing behavior, initial excitation should satisfy $R\mathrm{e}^{-\alpha d} \geq 1$, subsequently the gain coefficient will decrease with the increasing intensity inside the oscillation cavity, because of the gain saturation of an active medium, till a steady spasing is maintained. The theoretical spasing behavior resulted from the Fabry-Perot model is presented in Fig. 4, which shows a good agreement with the numerical simulations in Fig. 2. Some aspects about the theoretical results are discussed as follows:

Firstly, there is giant amplification, approximately 35.5 dB, in the transmittance as well as reflectance around the resonant frequency at a gain coefficient $\alpha_\mathrm{th} = -4.0\times10^4$ cm$^{-1}$. It is noticed that the equally enhanced reflection and transmission, demonstrated by the numerical simulation in Fig. 2, can be explained in the viewpoint of the Fabry-Perot effect.

Secondly, the spasing maximum is theoretically located at about $\alpha_\mathrm{th} = -4.0\times10^4$ cm$^{-1}$, far larger than the corresponding $\alpha_0 = -2.35\times10^3$ cm$^{-1}$ for the numerically expected spasing maximum. This is because that the slab is roughly equivalent in terms of the complex index of refraction $\tilde{n}$, and thus essentially it is a Fabry-Perot *laser* cavity, but never a *spaser* cavity with strong localized electromagnetic energy in gaps of metallic elements. Consequently, the equivalent slab cannot exhibit the advantage of an active surface plasmon resonator (namely, giant amplification at moderate gain level). As a matter of fact, an active medium in the gap of metals is usually enough to take sufficient enhancement, as long as strong electromagnetic energy in the form of plasmonic resonance is accumulated in the gap.[12,26] Qualitatively, according to the output intensity enhancement defined by Lambert law $I_{out} = I_{in} e^{-\alpha d}$, to reach the same enhanced $I_{out}$, the gain requirement in an active metamaterial reduces since inside the metamaterial $I_{local} \gg I_{in}$ because of the localized resonance rather than the gain effect. In another viewpoint,[11,12] it is considered that the effective gain coefficient of an active metamaterial could be substantially larger than that of the embedded bulk active medium itself due to the pronounced localized field in metamaterials. Anyway, the discrepancy in $\alpha_{th}$ and $\alpha_0$ for the spasing maximum can be well understood, and it implies that the localized plasmonic field in metallic metamaterials is responsible for the small gain requirement for giant enhancement.

Thirdly, the theoretical result in Fig. 4(b) confirms in further that the spasing intensity (i.e., the resonant transmittance and reflectance) does not increase monotonically with the gain coefficient, in consistent with the earlier numerical simulation [Fig. 2(b)]. According to the theoretical analysis of the Fabry-Perot model, a decreasing intensity of the spasing with increasing gain coefficient is resulted from the fact that the reflectance $R$ will approach to 1 when $\alpha$ (in proportion to $\kappa$ for given frequency) is overlarge and thus few input field can be coupled into (or out of) the Fabry-Perot cavity. It is worth of mentioning that the transmission reduction above a certain critical gain coefficient, though in accordance with literatures,[7,27] is possibly an unrealistic consequence of the time-independent solution, because physically there may be no way to obtain such overlarge values of gain coefficient in a steady state. Studies on this counterintuitive discrepancy have been conducted by solving time-dependent equations.[28]

Lastly, according to the theoretical result shown in Fig. 4(b), a dip of the reflected spasing should be emerged at $\alpha_{th} = 0 \text{ cm}^{-1}$, while in Fig. 2(b) this dip is numerically located at $\alpha_0 = -1.6 \times 10^3 \text{ cm}^{-1}$. This is because $\alpha_0$ in our numerical simulations is only the gain value for the active interlayer, while the net gain coefficient for the whole fishnet metamaterial should include the losses of the metallic bilayer. In other words, we can roughly evaluate the metallic loss (absorption coefficient) is $1.6 \times 10^3 \text{ cm}^{-1}$ so that the effective gain coefficient including the active layer and the metallic elements in the fishnet metamaterial is $0 \text{ cm}^{-1}$ at the reflected spasing dip.

### C. Directed emission of the spasing metamaterial

In spite of the fact that the spasing behavior in the active metamaterial was proposed as an analogy to the lasing effect and has been theorized in the frame of laser physics, spasing directivity has not been explored so far. Therefore, it should be of interest to investigate the emission directions of the spasing in the active fishnet metamaterial. As shown in Fig. 5(a), Instead of the normal incidence of a polarized plane wave used in simulations of the transmittance and reflectance (Fig. 2), we use a line source with an *x*-oriented current, randomly placed at $y = -2\ \mu$m and $z = -2\ \mu$m with respect to the center of the fishnet unit, to visualize the spasing directions in an explicit way.

According to the simulated result presented in Fig. 5(a), the transmitted as well as the reflected spasing has a strong intensity far more than the incident cylindrical wave does, and both of the transmission and reflection are preferentially directed along the normal of the fishnet layer (i.e., *z*-axis), approximately with equal enhancement. It should be emphasized that the radiation directivity is dependent on the resonant mode of the cavity. As for the spasing response in the fishnet metamaterial, it originates from the magnetic plasmon resonance with the current distribution shown in Figs. 5(b) and 5(c), where the antiparallel currents on the inner cavity surfaces of the opposite metallic layers witness the strong magnetic plasmon oscillation localized in the cavity. It is worthy to note that the subwavelength cavity thickness (about $\lambda_0 / 16$) is an attracting characteristic in cavity miniaturization.[29]

## IV. SUMMARY

A lasing spaser was recently demonstrated to have the ability of realizing a giant magnitude enhancement of transmission. In this work, to characterize the spasing properties from active metamaterials, a Fabry-Perot cavity slab with Lorentz dispersion in the index of refraction is studied to model the giant transmission as well as reflection. The analytical results from the equivalent slab show good agreement with the numerical simulations in aspects of spasing intensity and its gain dependence, except that the surface plasmon localization in the fishnet plasmonic resonator can not be covered by the equivalent model. It should be emphasized that, although the transmission and reflection in the fishnet spasing metamaterial can be interpreted in terms of a Fabry-Perot model as a laser is usually treated, an actual spaser cavity should be different from a laser cavity in that localized surface plasmon amplification plays an important role in the spasing behavior, especially for the small gain requirement. In addition, the subwavelength cavity thickness is distinguished from a conventional Fabry-Perot cavity.

## ACKNOWLEDGMENTS

This work was supported by the National Natural Science Foundation of China (Nos. 10604029, 10704036, 10874081, and 10904012).

[1]D. R. Smith, J. B. Pendry, and M. C. K. Wiltshire, Science **305**, 788 (2004).

[2]J. Valentine, S. Zhang, T. Zentgraf, E. Ulin-Avila, D. A. Genov, G. Bartal, and X. Zhang, Nature **455**, 376 (2008).

[3]J. B. Pendry, Phys. Rev. Lett. **85**, 3966 (2000).

[4]X. Zhang and Z. Liu, Nat. Materials **7**, 435 (2008).

[5]J. B. Pendry, D. Schurig, and D. R. Smith, Science **312**, 1780 (2006).

[6]R. Liu, C. Ji, J. J. Mock, J. Y. Chin, T. J. Cui, and D. R. Smith, Science **323**, 366 (2009).

[7]N. I. Zheludev, S. L. Prosvirnin, N. Papasimakis, and V. A. Fedotov, Nat. Photonics **2**, 351 (2008).

[8]M. A. Noginov, G. Zhu, A. M. Belgrave, R. Bakker, V. M. Shalaev, E. E. Narimanov, S. Stout, E. Herz, T. Suteewong, and U. Wiesner, Nature **460**, 1110 (2009).

[9]Z. H. Zhu, H. Liu, S. M. Wang, T. Li, J. X. Cao, W. M. Ye, X. D. Yuan, and S. N. Zhu, Appl. Phys. Lett. **94**, 103106 (2009).

[10]N. Papasimakis, V. A. Fedotov, Y. H. Fu, D. P. Tsai, and N. I. Zheludev, Phys. Rev. B **80**, 041102(R) (2009).

[11]M. Wegener, J. L. Garcia-Pomar, C. M. Soukoulis, N. Meinzer, M. Ruther, and S. Linden, Opt. Express **16**, 19785 (2008).

[12]A. Fang, Th. Koschny, M. Wegener, and C. M. Soukoulis, Phys. Rev. B **79**, 241104(R) (2009).

[13]S. Zhang, W. Fan, K. J. Malloy, S. R. J. Brueck, N. C. Panoiu, and R. M. Osgood, Opt. Express **13**, 4922 (2005).

[14]M. Kafesaki, I. Tsiapa, N. Katsarakis, Th. Koschny, C. M. Soukoulis, and E. N. Economou, Phys. Rev. B **75**, 235114 (2007).

[15]K. Aydin, Z. Li, L. Sahin, and E. Ozbay, Opt. Express **16**, 8835 (2008).

[16]T. Li, S.-M. Wang, H. Liu, J.-Q. Li, F.-M. Wang, S.-N. Zhu, and X. Zhang, J. Appl. Phys. **103**, 023104 (2008).

[17]Z. G. Dong, M. X. Xu, S. Y. Lei, H. Liu, T. Li, F. M. Wang, and S. N. Zhu, Appl. Phys. Lett. **92**, 064101 (2008).

[18]G. Dolling, C. Enrich, M. Wegener, C. M. Soukoulis, and S. Linden, Opt. Lett. **31**, 1800 (2006).

[19]E. Plum, V. A. Fedotov, P. Kuo, D. P. Tsai, and N. I. Zheludev, Opt. Express **17**, 8548 (2009).

[20]F. M. Wang, H. Liu, T. Li, Z. G. Dong, S. N. Zhu, and X. Zhang, Phys. Rev. E **75**, 016604 (2007).

[21]Z. G. Dong, H. Liu, T. Li, Z. H. Zhu, S. M. Wang, J. X. Cao, S. N. Zhu, and X. Zhang, Opt. Express **16**, 20974 (2008).

[22]D. J. Bergman and M. I. Stockman, Phys. Rev. Lett. **90**, 027402 (2003).

[23]M. I. Stockman, Nat. Phtonics **2**, 327 (2008).

[24]N. Hodgson and H. Weber, *Laser Resonators and Beam Propagation* (Springer, New York, 2005).

[25]X. Jiang, Q. Li, and C. M. Soukoulis, Phys. Rev. B **59**, R9007 (1999).

[26]J. B. Pendry, A. J. Holden, D. J. Robbins, and W. J. Stewart, IEEE Trans. Microwave Theory Tech. **47**, 2075 (1999).

[27]E. Poutrina and D. R. Smith, in *Proceedings of Conference on International Quantum Electronics*, *Maryland*, *2009*, OSA Technical Digest (CD) (Optical Society of America, Washington, 2009), Paper No. ITuD6.

[28]H. Bahlouli, A. D. Alhaidari, A. Al Zahrani, and E. N. Economou, Phys. Rev. B **72**, 094304 (2005).

[29]L. Zhou, H. Li, Y. Qin, Z. Wei, and C. T. Chan, Appl. Phys. Lett. **86**, 101101 (2005).

**Figure captions:**

FIG. 1. (Color online) The schematic illustration of the spasing metamaterial with fishnet structure. (a) The excitation configuration. (b) Scale symbols for a unit cell.

FIG. 2. (Color online) Simulation results from the spasing fishnet metamaterial. (a) Spectra of the transmission and reflection. (b) Gain dependence of the transmitted and reflected spasing intensities.

FIG. 3. (Color online) Ray diagram of a Fabry-Perot slab with the amplitudes of multiple transmissions and reflections indicated. The amplitude factor $f' = R e^{-i\varphi} e^{-\alpha d}$ and the incident angle $i_1 = 0$ for a normal incidence.

FIG. 4. (Color online) Theoretical results from the equivalent Fabry-Perot slab. (a) Spectra of the transmission and reflection. (b) Gain dependence of the transmitted and reflected spasing intensities.

FIG. 5. (Color online) (a) The directed spasing field distribution from the active fishnet unit located at the *zy*-face center of the cylinder with $5\,\mu$m radius. The excitation is an *x*-oriented line source at $y = -2\,\mu$m and $z = -2\,\mu$m with respect to the fishnet center. For the vacuum cylinder, the bases are perfect electric boundaries and the lateral surface is a radiation boundary. (b and c) The antiparallel current distributions on the opposite inner surfaces of the cavity.

FIG. 1. Z. G. Dong et al.

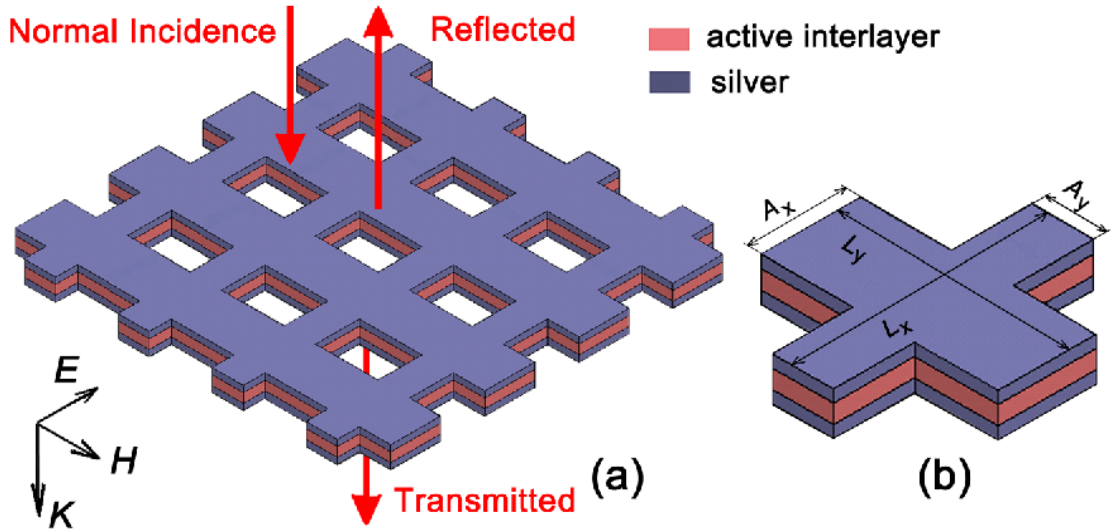

FIG. 2. Z. G. Dong et al.

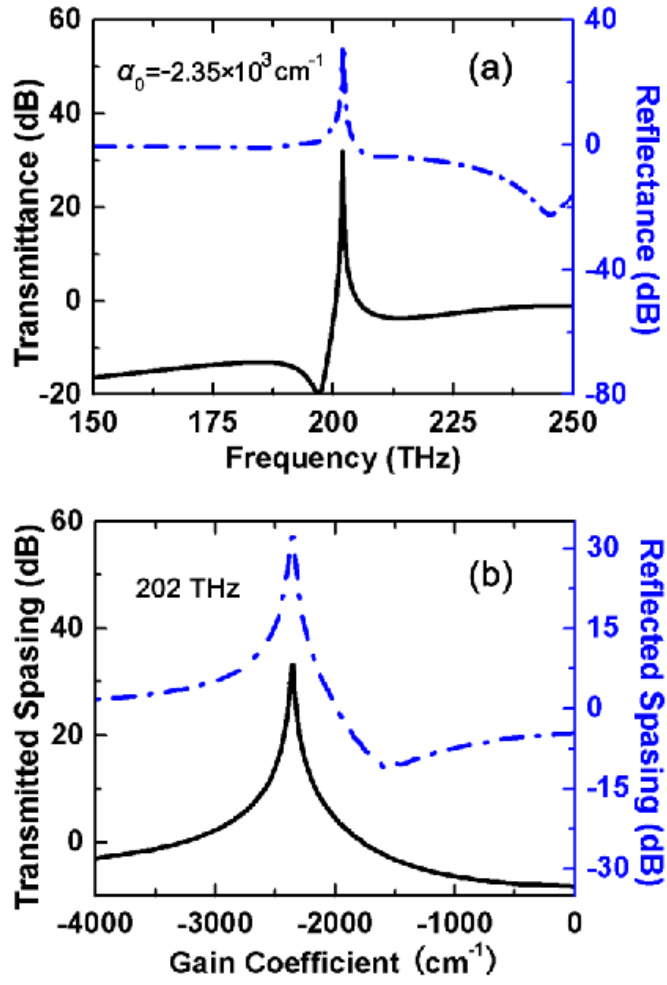

FIG. 3. Z. G. Dong et al.

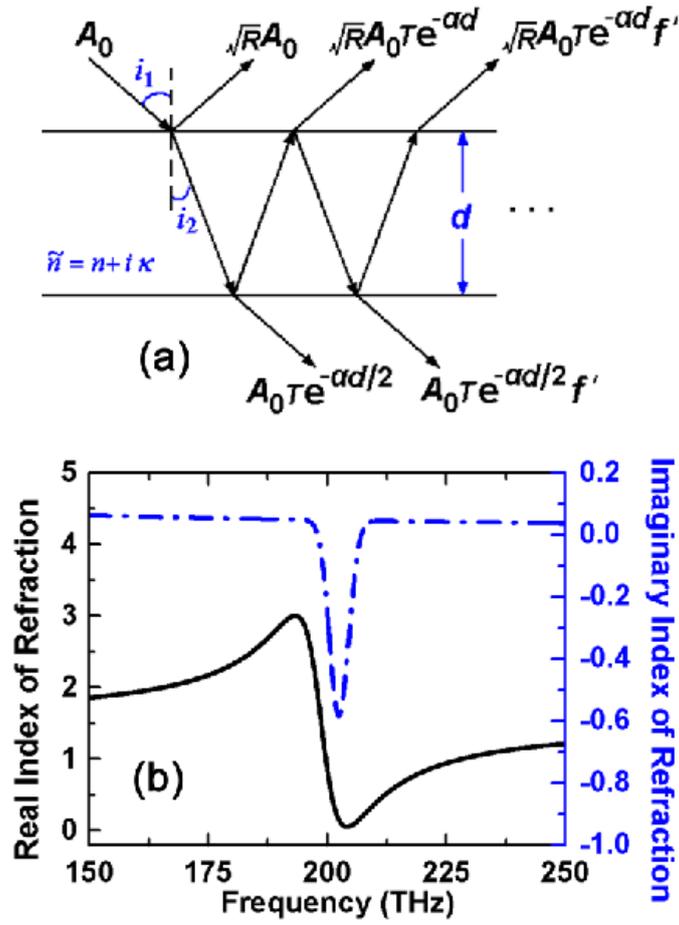

FIG. 4. Z. G. Dong et al.

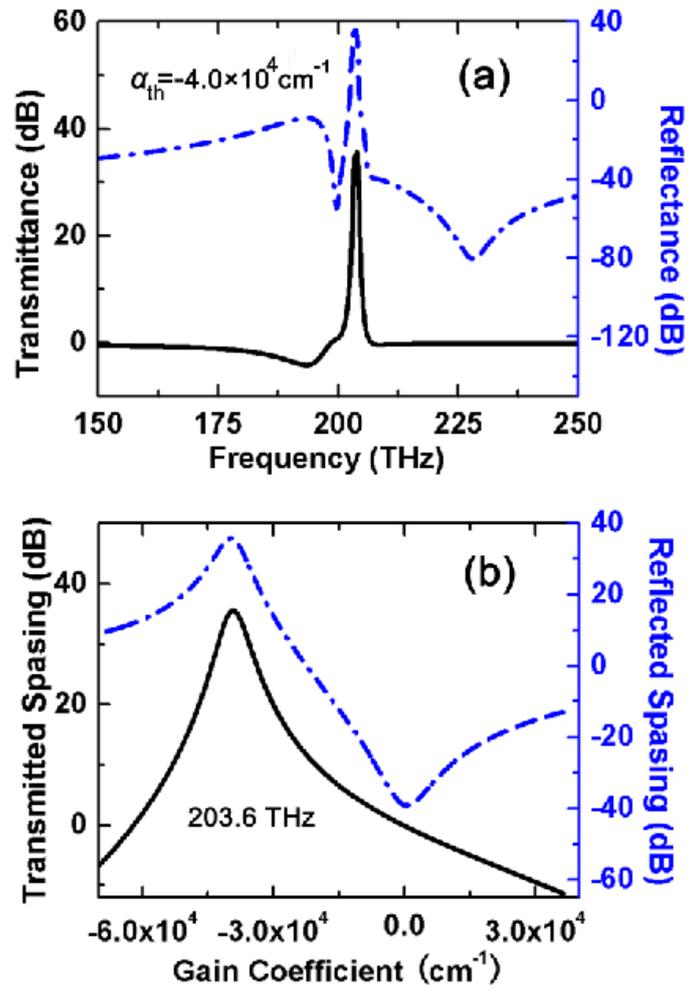
19

FIG. 5. Z. G. Dong et al.

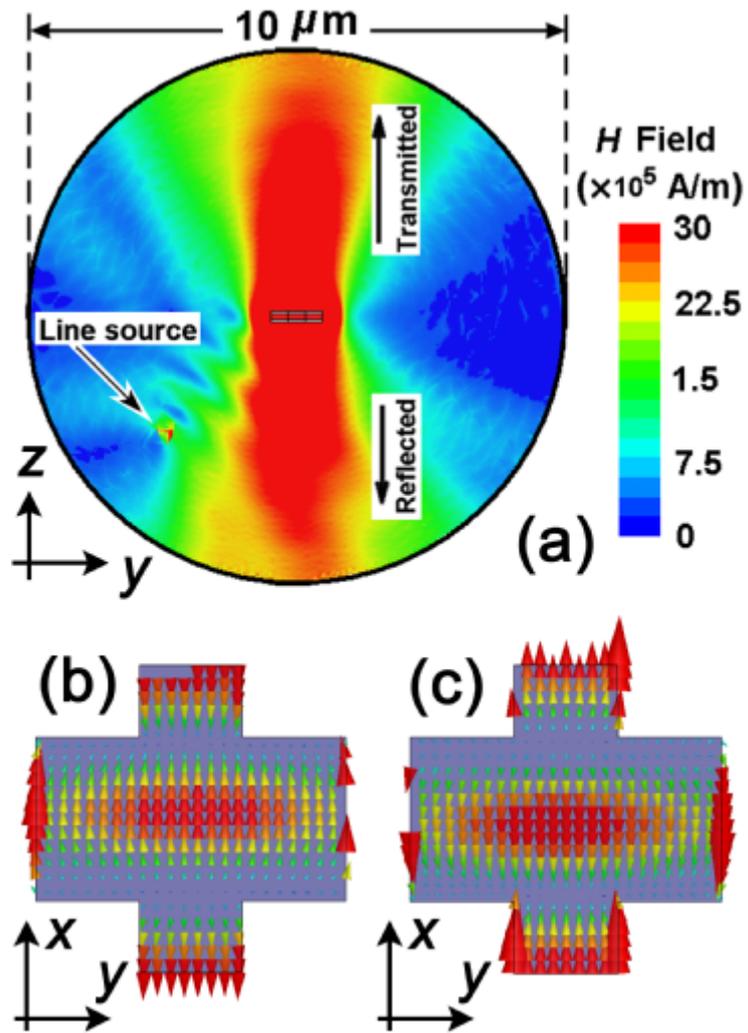
20